\documentclass{article}
\usepackage{ijcai23}

\usepackage{times}
\usepackage{soul}
\usepackage{url}
\usepackage[hidelinks]{hyperref}
\usepackage[utf8]{inputenc}
\usepackage[small]{caption}
\usepackage{graphicx}
\usepackage{amsmath}
\usepackage{amsthm}
\usepackage{booktabs}
\usepackage[switch]{lineno}

\usepackage{xcolor}
\newcommand{\newcite}[1]{\citeauthor{#1}~\shortcite{#1}}
\usepackage{multirow}
\usepackage{amssymb}
\usepackage{booktabs}
\usepackage{algpseudocode}

\usepackage[ruled,linesnumbered,vlined]{algorithm2e}
\SetKwFunction{fm}{C\_SCL}
\SetKwFunction{fcc}{\textcolor{magenta}{cur\_cri}}
\SetKwFunction{fcl}{\textcolor{magenta}{cur\_lev}}
\SetKwFunction{fcs}{\textcolor{magenta}{cur\_sche}}

\newcommand{\CalS}{\mathcal{S}}
\newcommand{\CalL}{\mathcal{L}}

\newcommand{\Vh}{\boldsymbol{\mathit{h}}}

\newcommand{\Ve}{\boldsymbol{\mathit{e}}}

\newcommand{\SetE}{\mathbb{E}}
\newcommand{\SetS}{\mathbb{S}}
\newcommand{\SetP}{\mathbb{P}}
\newcommand{\SetR}{\mathbb{R}}

\newcommand{\Mx}{\boldsymbol{\mathit{X}}} 
\newcommand{\My}{\boldsymbol{\mathit{Y}}} 
\newcommand{\Me}{\boldsymbol{\mathit{E}}} 

\title{Semantic-aware Contrastive Learning for Electroencephalography-to-Text Generation with Curriculum Learning}

\author{
	Xiachong Feng
	\and
	Xiaocheng Feng 
	\and
	Bing Qin
	\affiliations
	Harbin Institute of Technology, China
	\emails
\{xiachongfeng, xcfeng, bqin\}@ir.hit.edu.cn
}

\begin{document}

\maketitle

\begin{abstract}
Electroencephalography-to-Text generation  (EEG-to-Text), which aims to directly generate natural text from EEG signals has drawn increasing attention in recent years due to the enormous potential for Brain-computer interfaces (BCIs).
However, the remarkable discrepancy between the subject-dependent EEG representation and the semantic-dependent text representation poses a great challenge to this task.
To mitigate this challenge, we devise a \textbf{C}urriculum \textbf{S}emantic-aware \textbf{C}ontrastive \textbf{L}earning strategy (\textsc{C-SCL}), which effectively re-calibrates the subject-dependent EEG representation to the semantic-dependent EEG representation, thus reducing the discrepancy.
Specifically, our \textsc{C-SCL} pulls semantically similar EEG representations together while pushing apart dissimilar ones.
Besides, in order to introduce more meaningful contrastive pairs, we carefully employ curriculum learning to not only craft meaningful contrastive pairs but also make the learning progressively.
We conduct extensive experiments on the ZuCo benchmark and our method combined with diverse models and architectures shows stable improvements across three types of metrics while achieving the new state-of-the-art.
Further investigation proves not only its superiority in both the single-subject and low-resource settings but also its robust generalizability in the zero-shot setting\footnote{Our codes and models will be made public.}.
\end{abstract}

\section{Introduction}

\begin{figure}[t]
    \centering
    \includegraphics[scale=0.50]{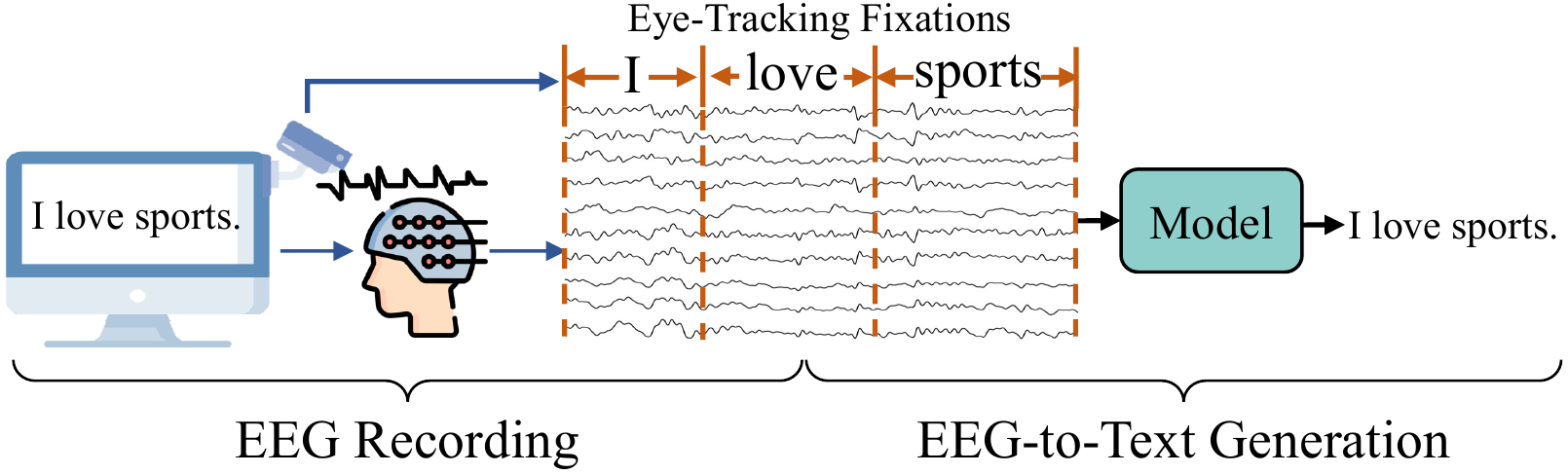}
    \caption{Illustration of the EEG-to-Text generation task. The left part shows the EEG recording process, in which one subject reads a sentence on the screen while recording its EEG signals. Concurrently, the eye-tracking device permits defining exact word boundaries via fixations. Given the recorded EEG signals, the task aims to generate the sentence that stimulates those EEG signals.}
    \label{fig:intro}
\end{figure}

Devastating neurological conditions such as spinal cord injuries or neuromuscular disorders can suddenly lead to people losing their ability to communicate \cite{stanger1996demographics,claassen2019detection}. Such patients may still have intact language and cognitive skills, but injuries might hinder them from expressing themselves \cite{felgoise2016verbal}.
Fortunately, Brain-computer interfaces (BCIs) can restore language abilities to such patients by decoding neural activities into the natural language (Brain-to-Text), which can drastically improve their quality of life \cite{brumberg2018brain}. 
To pursue this goal, various Brain-to-Text works are proposed, building upon either \textit{invasive brain recordings}, such as electrocorticography (ECoG) \cite{anumanchipalli2019speech,makin2020machine,metzger2022generalizable}, or \textit{non-invasive brain recordings}, such as functional magnetic resonance imaging (fMRI) \cite{zou2022cross} and electroencephalography (EEG) \cite{wang2022open}. 
Amongst, EEG shows its superior benefits in portability and cost-effectiveness in real-world applications, thus EEG-to-Text generation gains a lot of research interest recently \cite{wang2022open,defossez2022decoding}.
Figure~\ref{fig:intro} depicts the task flow.

\begin{figure}[t]
    \centering
    \includegraphics[scale=0.42]{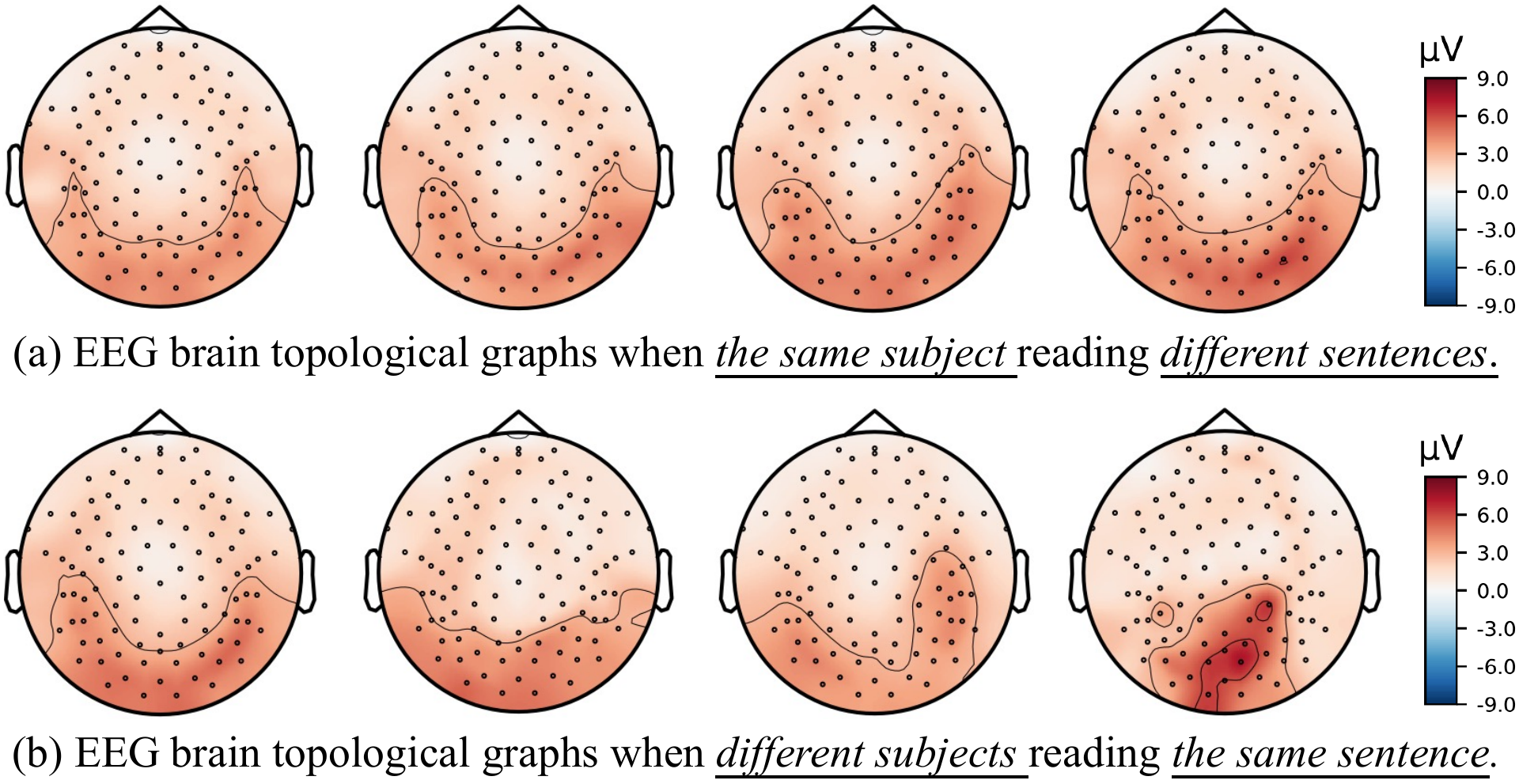}
    \caption{Brain topological graph of the sentence-level EEG representation (averaged word-level EEG representations). (a) Four topological graphs denote the EEG representations elicited by the same subject in response to four different sentences. (b) Four topological graphs describe the EEG representations elicited by four different subjects corresponding to the same sentence.}
    \label{fig:brain_topo}
\end{figure}

However, we claim that existing studies neglect the discrepancy between the \textit{subject-dependent EEG representation} and the \textit{semantic-dependent text representation}, which inevitably degrades EEG-to-Text model performance.
To explain why it becomes a crucial challenge for this task, we present brain topological graphs to intuitively visualize the discrepancy under two situations.
Firstly, as shown in Figure~\ref{fig:brain_topo}(a), EEG representations elicited by the same subject skewed towards being similar, no matter what the sentence stimulus is, demonstrating the same subject is prone to favour similar cognitive patterns in the face of different sentence stimuli.
Secondly, on the contrary, Figure~\ref{fig:brain_topo}(b) reveals that different subjects act variously even disparately in terms of the same sentence stimulus.
These observations are in line with findings in previous studies, including neuroscience \cite{adolphs2002neural} as well as some machine learning research areas, such as emotion classification \cite{shen2022contrastive} and visual recognition \cite{lee2022inter}.
On this account, such {subject-dependent EEG representation} negatively impacts the performance of EEG-to-Text model from two perspectives.
On the one hand, it introduces a ``many-to-one" generation problem (multiple EEG signals correspond to the same sentence), which is challenging for training current sequence-to-sequence generation models.
On the other hand, it largely hinders good cross-subject generalizability since transferring original {subject-dependent EEG representation} to unseen subjects is intractable.

To address this issue, we propose a novel \textbf{C}urriculum \textbf{S}emantic-aware \textbf{C}ontrastive \textbf{L}earning strategy (\textsc{C-SCL}), which can effectively re-calibrate the original {subject-dependent EEG representation} into our desirable {semantic-dependent EEG representation} so that it can be better adapted to the EEG-to-Text generation task.
In detail, the core part of our \textsc{C-SCL} is the  \textbf{S}emantic-aware \textbf{C}ontrastive \textbf{L}earning strategy (\textsc{SCL}), which aims to maximize the similarities of EEG representations across subjects \textit{w.r.t.} the identical sentence stimulus (positive pairs) while minimizing the similarities of EEG representations \textit{w.r.t.} the different sentence stimuli (negative pairs).
Note that the critical ingredient for successful contrastive learning is to construct hard positive and negative pairs. 
However, based on the random selection, we witness that nearly 45.93\% of total constructed contrastive pairs  already satisfy the final objective, where positive pairs are similar and negative pairs are dissimilar.
Therefore, we manufacture contrastive pairs in different difficulties by pre-computing similarities between numerical EEG signals (e.g., hard positive pairs initially have low similarity while hard negative pairs have high similarity) and drawing support from curriculum learning to not only introduce hard contrastive pairs but also enable a progressive learning process by learning from easy pairs to hard pairs.
With the integration of curriculum learning, we finalize our \textbf{C}urriculum \textbf{S}emantic-aware \textbf{C}ontrastive \textbf{L}earning strategy (\textsc{C-SCL}).

We conduct experiments on the ZuCo benchmark \cite{hollenstein2018zuco,hollenstein-etal-2020-zuco} and evaluate the generation performance via three types of metrics.
Experimental results show the effectiveness of our proposed method across various models and architectures.
Further investigation empirically shows its benefits in both the single-subject setting and low-resource settings as well as its robust generalizability in the zero-shot setting. In summary: (a) We take the first step to mitigating the challenge of the discrepancy between the subject-dependent EEG representation and the semantic-dependent text representation for the EEG-to-Text generation task; (b) We devise a curriculum semantic-aware contrastive learning strategy that succeeds at yielding the semantic-dependent EEG representation; (c) We conduct extensive experiments on the ZuCo benchmark that demonstrates the effectiveness of our method and its robustness and superior generalizability.

\section{Related Work}
\subsection{Brain-to-Text Generation}
Brain-to-Text generation task aims to generate the text given corresponding brain signals.
According to the classification criterion of vocabulary size, there are two series of related works: \textit{closed vocabulary} and \textit{open vocabulary} brain-to-text generation. 
The first line of works generates words in small closed vocabularies \cite{makin2020machine,moses2021neuroprosthesis}. 
For example, \newcite{moses2021neuroprosthesis} focus on a 50-word vocabulary.
While exhibiting promising generation accuracy and speed, expanding access to a larger vocabulary enables effective day-to-day communication.
Accordingly, \newcite{wang2022open} study the problem of open vocabulary EEG-to-Text decoding task by utilizing pre-trained language models (PLMs) \cite{Lewis2019BARTDS}. It brings two benefits: on the one hand, PLMs offer a large vocabulary, on the other hand, PLMs can serve as a bridge between brain signals and linguistic information \cite{millet2022toward}.
In our work, we focus on the open vocabulary EEG-to-Text generation due to the non-invasive nature and widespread application prospects of EEG-based BCIs. 
Specifically, we pay particular attention to the challenge of the discrepancy between the subject-dependent EEG representation and the semantic-dependent text representation for the EEG-to-Text generation task.

\subsection{Contrastive Learning}
Contrastive learning is a technique that aims to make the representation of a given anchor data to be similar to its positive pairs while being dissimilar to its negative pairs.
It shows promising results in computer vision \cite{hadsell2006dimensionality,chen2020simple,he2020momentum} and has gained popularity in natural language processing \cite{giorgi-etal-2021-declutr,gao2021simcse}.
After witnessing its superiority in the above areas, contrastive learning is attracting the attention of neuroscientists and has been applied to several EEG-based classification tasks \cite{pmlr-v136-mohsenvand20a,Cheng2020SubjectAwareCL,defossez2022decoding,shen2022contrastive,lee2022inter}.
More recently, \newcite{shen2022contrastive} propose a contrastive learning method to tackle the cross-subject emotion recognition problem. \newcite{defossez2022decoding} devise a contrastive learning objective to align representations of brain signals and natural speech.
In our work, we devise a novel curriculum semantic-aware contrastive learning strategy, aiming to learn semantic-dependent EEG representations, which effectively reduce the discrepancy between the EEG and text representations. 

\begin{table}[t]
\centering
        \begin{tabular}{l|ccc}
            \toprule
            & \textbf{Train} & \textbf{Valid} & \textbf{Test}  \\
            \midrule
            \# pairs &14567  &1811  &1821   \\
            \# unique\_sent &1061  &173  &146   \\
            \# subject &30  &30  &30   \\
            avg.words &19.89  &18.80  &19.23 \\
            \bottomrule
        \end{tabular}
\caption{Statistics for the ZuCo benchmark. ``\# pairs" means the number of EEG-text pairs, ``\# unique\_sent" represents the number of unique sentences, ``\# subject" denotes the number of subjects and ``avg.words" means the average number of words of sentences.}
\label{tab:dataset}
\end{table}

\section{Preliminaries}
In this section, we first describe the task formulation and then introduce the ZuCo benchmark.

\subsection{Task Formulation}\label{sec:task}
Given a sequence of word-level EEG features $\Me$, EEG-to-Text generation task aims at producing a sentence $\CalS$ via a model $\theta$, where $\Me$ consists of $|\Me|$ features $[{\Ve}_1, {\Ve}_2, ..., {\Ve}_{|\Me|}]$ and $\CalS$ consists of $|\CalS|$ tokens $[s_1, s_2, ..., s_{|\CalS|}]$.
$\Ve \in {\SetR}^n$ symbolizes a word-level EEG feature vector and $\theta$ denotes the parameters of a sequence-to-sequence model.
Each sequence of EEG features $\Me$ is associated with a subject $p_i \in \SetP$, $\SetP$ being a set of subjects.
During the training phase, EEG-Text pairs come from various subjects and the learning objective.
At the test phase, sentences are totally unseen. Besides, the train, valid and test sets maintain the same set of subjects $\SetP$.

\subsection{ZuCo Benchmark}
We use the ZuCo dataset, which is a corpus of EEG signals and eye-tracking data during natural reading. 
The reading materials are collected from movie reviews and Wikipedia articles.
Specifically, following \newcite{wang2022open}, we utilize the combination of both ZuCo \cite{hollenstein2018zuco} and ZuCo 2.0 \cite{hollenstein-etal-2020-zuco} to form our final ZuCo benchmark.
For each EEG-text pair in the dataset, EEG signals are composed of a sequence of word-level EEG features $\Me$. 
For each word-level feature $\Ve$, 8 frequency bands are recorded and denoted as the following: theta1 (4-6Hz), theta2 (6.5–8Hz), alpha1 (8.5–10Hz), alpha2 (10.5–13Hz), beta1 (13.5–18Hz) beta2 (18.5–30Hz) and gamma1 (30.5–40Hz) and gamma2 (40–49.5Hz).
Each band of the feature has a fixed dimension of 105.
We concatenate all 8 bands of features to construct the final word-level feature vector with a dimension of 840 ($\Ve\in{\SetR}^{840}$). 
Additionally, all features are Z-scored as done by \newcite{willett2021high}.
We further split the dataset into train, valid and test (80\%,10\%,10\%) parts following \newcite{wang2022open}.
Note that each part of the dataset maintains the same subject set with no overlapping sentences.
Table \ref{tab:dataset} shows the statistics of the ZuCo benchmark\footnote{We omit EGG signals that contain \textit{NaN} values following \newcite{wang2022open}. Therefore, different subjects may associate with different sentence sets.}.

\begin{figure}[t]
    \centering
    \includegraphics[scale=0.58]{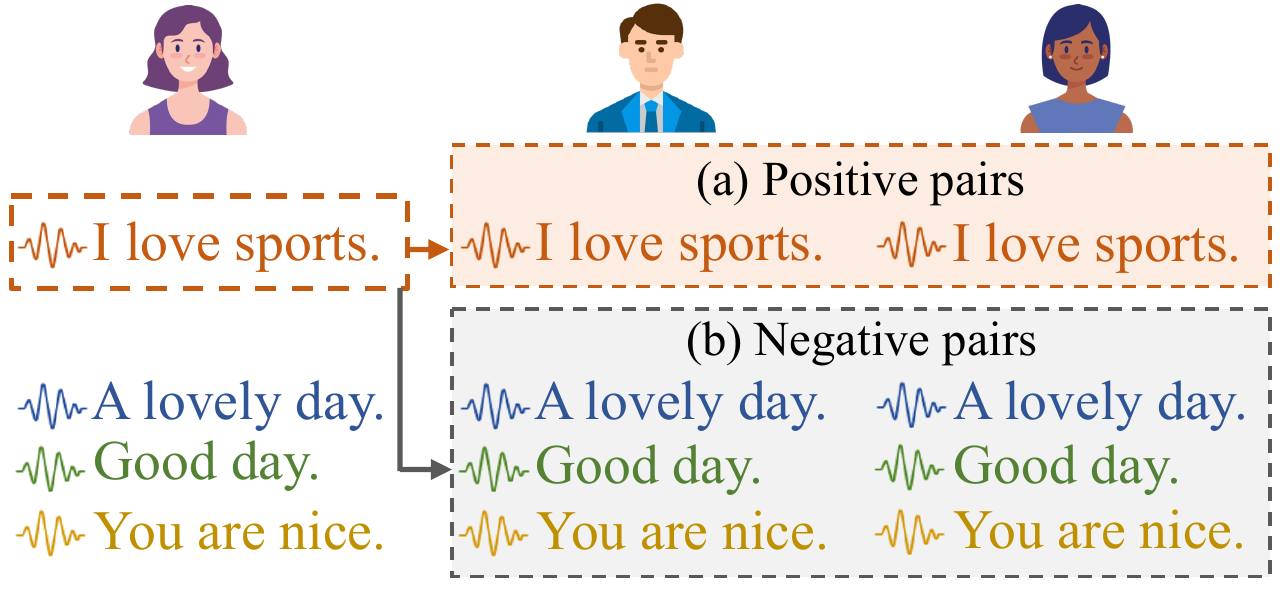}
    \caption{Illustration of our semantic-aware contrastive learning strategy (\textsc{SCL}). (a) Positive pairs derive from EEG signals corresponding to the same sentence elicited by different subjects. In contrast, (b) Negative pairs come from EEG signals elicited by different subjects corresponding to different sentences.}
    \label{fig:cl}
\end{figure}

\section{Method}
In this section, we thoroughly introduce our curriculum semantic-aware contrastive learning strategy (\textsc{C-SCL}) step by step, including (1) semantic-aware contrastive learning, (2) curriculum learning, (3) the backbone model \textsc{BrainTranslator} and (4) the overall learning procedure.

\subsection{Semantic-aware Contrastive Learning}

\noindent \paragraph{Motivation.} The critical ingredient of training a superior model for EEG-to-Text generation is reducing the discrepancy between the \textit{subject-dependent EEG representation} and the \textit{semantic-dependent text representation}.
To this end, we draw support from contrastive learning \cite{hadsell2006dimensionality}, which is skilled at re-calibrating the representation space, and propose our semantic-aware contrastive learning strategy (\textsc{SCL}) by pulling semantically similar EEG representations together (positive pairs) and pushing apart dissimilar ones (negative pairs).
Note that we use intrinsic semantics behind the EEG signals as a criterion to guide the surface EEG contrastive learning, which implicitly aligns two modalities, thus achieving the \textit{semantic-dependent EEG representation}.

\noindent \paragraph{Positive Pairs.}
One important question in contrastive learning is how to construct positive pairs $(\Me_i, \Me_i^+)$.
Towards achieving our goal of learning {semantic-dependent EEG representations}, given an anchor EEG representation $\Me_i$ with its corresponding sentence $\CalS_i$, we randomly choose one EEG $\Me_i^+$ from the positive set $\SetE_i^+$, in which all EEG signals are correspond to the same sentence stimulus $\CalS_i$ across different subjects, as shown in Figure~\ref{fig:cl}(a).
Such positive pairs will promote clustering 
 of semantically similar EEG signals.

\noindent \paragraph{Negative Pairs.}
Practically speaking, original in-batch negatives insufficiently provide weak supervision for contrastive learning. To alleviate this problem, \newcite{gao2021simcse} verify that introducing specially designed negative pairs can stimulate the learning process.
Inspired by this conclusion, given the anchor EEG representation $\Me_i$ elicited by $p_i$ with its corresponding sentence $\CalS_i$, we construct the negative pair $(\Me_i, \Me_i^-)$, where $\Me_i^-$ satisfies two conditions\footnote{In our preliminary experiments, we consider both two conditions and only the first condition, the results show that considering both conditions can achieve better results.}: (1) $\Me_i^-$ corresponds to sentences except for $\CalS_i$ and (2) $\Me_i^-$ is elicited by subjects except for $p_i$.
All $\Me_i^-$ that satisfy both conditions form the negative set $\SetE_i^-$, as shown in Figure~\ref{fig:cl}(b).

\subsection{Curriculum Learning}

\noindent \paragraph{Motivation.}
The key point for effective contrastive learning is to construct hard contrastive pairs.
However, after the examination of our \textsc{SCL}, we witness that nearly 45.93\% of total constructed contrastive pairs already satisfy the final objective\footnote{We run \textsc{SCL} for 10 epochs on the ZuCo train set, resulting in 145670 (14567$\times$10) contrastive triples, in which 66906 triples already satisfy the final objective, leading to $\frac{66906}{145670}=45.93\%$.}.
To overcome this problem, we employ curriculum learning to not only introduce hard contrastive pairs but also ensure the model learning efficiency, thus finalizing our \textbf{C}urriculum \textbf{S}emantic-aware \textbf{C}ontrastive \textbf{L}earning strategy (\textsc{C-SCL}).
Compared with \textsc{SCL} that randomly selects positive sample and negative sample from $\SetE_i^+$ and $\SetE_i^-$ respectively, \textsc{C-SCL} selects samples in an easy-to-hard order.

\noindent \paragraph{Curriculum Criterion.}
\textit{How to determine the ordering?} 
Recall that our goal is to introduce hard contrastive pairs, where positive pairs are initially far away from each other while negative pairs are oppositely similar. 
Therefore, we pre-calculate the cosine similarity between two EEG representations and craft contrastive pairs of varying difficulties by taking the pre-calculated similarity into consideration. 
Specifically, given an anchor EEG representation $\Me_i$, for positive pair construction, we calculate similarities between the $\Me_i$ and all $\Me_i^+ \in \SetE_i^+$ and then sort the $\SetE_i^+$ in the descending order, resulting in $\grave{\SetE}_i^+$.
On the contrary, for negative set $\SetE_i^-$, we sort it in the ascending order and attain $\acute{\SetE}_i^-$.
Both hard positive and negative samples \textit{w.r.t.} the anchor $\Me_i$  are located at the end of the $\grave{\SetE}_i^+$ and $\acute{\SetE}_i^-$, respectively. 
In other words, samples in the $\grave{\SetE}_i^+$ and $\acute{\SetE}_i^-$ are now in an easy-to-hard order.

\noindent \paragraph{Curriculum Level.}
\textit{What are the curriculum levels?} 
We conduct preliminary experiments by setting up the number of curriculum levels from 2 to 5 and finally decide to split the $\grave{\SetE}_i^+$ and $\acute{\SetE}_i^-$ into 3 levels due to their better performance.
In detail, we split the sorted $\grave{\SetE}_i^+$ into three consecutive sets, including $[{\SetE}_i^{\texttt{easy}+}, {\SetE}_i^{\texttt{medium}+}, {\SetE}_i^{\texttt{hard}+}]$ and $\acute{\SetE}_i^-$ into $[{\SetE}_i^{\texttt{easy}-}, {\SetE}_i^{\texttt{medium}-}, {\SetE}_i^{\texttt{hard}-}]$\footnote{Example contrastive pairs of different difficulties are shown in the supplementary file.}.

\noindent \paragraph{Curriculum Scheduler.}
\textit{When to update the curriculum?} 
We adopt an One-Pass scheduler with a linear pace \cite{bengio2009curriculum} to progressively train the model in an easy-to-hard order, 
For example, when reaching to the hard level, given an anchor EEG $\Me_i$, we select positive sample and negative sample from ${\SetE}_i^{\texttt{hard}+}$ and ${\SetE}_i^{\texttt{hard}-}$, respectively.

\begin{algorithm}[t]
\small
	\KwIn{EEG $\Me_i$ with its corresponding subject $p_i$ and sentence $\CalS_i$; 
		a dict $f_s: \CalS_i \rightarrow \SetE_{\CalS_i}$ maps $\CalS_i$ to a set of EEG signals $\SetE_{\CalS_i}$;
            a dict $f_p: p_i \rightarrow \SetE_{p_i}$ maps $p_i$ to a set of EEG signals $\SetE_{p_i}$;
		a set of all sentences $\SetS$;
		current curriculum level $\texttt{curr\_level}$;

	}
	\KwOut{a contrastive triple $(\Me_i, \Me_i^+, \Me_i^-)$.}

	\SetKwProg{Fn}{Function}{:}{} 

 	\Fn{\fm{$\Me_i$, \texttt{curr\_level}}}{
		\tcp{\textcolor{blue}{{positive sample}}}
		$\SetE_i^+$ = ${f_s}(\CalS_i)\backslash \Me_i$\;
		$\grave{\SetE}_i^+$ = \textcolor{magenta}{\texttt{cur\_cri}}($\Me_i$, $\SetE_i^+$, \texttt{descending})\;
  		\texttt{curriculums} = \textcolor{magenta}{\texttt{cur\_lev}}($\grave{\SetE}_i^+$)\;
		$\Me_i^+$ = $\textcolor{magenta}{\texttt{cur\_sche}}(\texttt{curriculums}, \texttt{curr\_level})$\;

   	\tcp{\textcolor{blue}{{negative sample}}}

		$\SetE_i^- = {f_s}(\SetS \backslash \CalS_i)-{f_p}(p_i)$\;
		$\acute{\SetE}_i^-$ = \textcolor{magenta}{\texttt{cur\_cri}}($\Me_i$, $\SetE_i^-$, \texttt{ascending})\;
            \texttt{curriculums} = \textcolor{magenta}{\texttt{cur\_lev}}($\acute{\SetE}_i^-$)\;
        $\Me_i^-$ = $\textcolor{magenta}{\texttt{cur\_sche}}(\texttt{curriculums}, \texttt{curr\_level})$\; 
			
		\KwRet $(\Me_i, \Me_i^+, \Me_i^-)$\;
	}
		            	
 	\tcp{\textcolor{blue}{{curriculum criterion}}}
	\SetKwProg{Fn}{Function}{:}{}
	\Fn{\fcc{$\Me$, $\SetE$, \texttt{order}}}{
		\texttt{sims} = \texttt{list()}\;
		\For{$\Me_j \in \SetE$}{
			$\texttt{sim}_{\texttt{j}} = \texttt{cosine\_similarity}(\Me , \Me_j)$\;
			$\texttt{sims.append}(\texttt{sim}_{\texttt{j}})$
		}
  		\texttt{indices} = \texttt{sims.sort(order)}\;
		\KwRet $\SetE[\texttt{indices}]$\;
	}

	\tcp{\textcolor{blue}{{curriculum level}}}
	\SetKwProg{Fn}{Function}{:}{}
	\Fn{\fcl{$\SetE$}}{
            $[{\SetE}^{\texttt{easy}}, {\SetE}^{\texttt{medium}}, {\SetE}^{\texttt{hard}}]$ = \texttt{split}($\SetE$)\;
		\KwRet $[{\SetE}^{\texttt{easy}}, {\SetE}^{\texttt{medium}}, {\SetE}^{\texttt{hard}}]$\;	
	}
		
	\tcp{\textcolor{blue}{{curriculum scheduler}}}
	\SetKwProg{Fn}{Function}{:}{}
	\Fn{\fcs{$\texttt{curriculums}$, $\texttt{curr\_level}$}}{
		${\SetE}^{\texttt{select}}$ = \texttt{select}(\texttt{curriculums}, $\texttt{curr\_level}$)\;
          $\Me$ = \texttt{random\_select}(${\SetE}^{\texttt{select}}$)\;

			\KwRet $\Me$\;	
	}	
	\caption{Curriculum Semantic-aware Contrastive Learning
		}\label{algo:cl}
\end{algorithm}

\begin{table*}[t]
\small
\centering
        \begin{tabular}{lcccccccc}
            \toprule
            &  \multicolumn{3}{c}{\textbf{ROUGE(\%)↑}} & \multicolumn{4}{c}{\textbf{BLEU(\%)↑}} & \multirow{2}{*}{\textbf{WER(\%)↓}} \\
            \textbf{Model} & \textbf{R-1} & \textbf{R-2} & \textbf{R-L} & \textbf{B-1} & \textbf{B-2} & \textbf{B-3} & \textbf{B-4} & \\
            \midrule
            \textsc{BrainBART-Large} \cite{wang2022open} &37.85 &18.83 &35.92 &34.79 &24.38 &19.58 &17.02 & 70.31  \\
            \textsc{BrainBART-Large} (w/ \textsc{SCL}) &38.71 &20.05 &36.73 &35.65 &25.74 &\textbf{21.31} &\textbf{18.96} & 69.12  \\
            \textsc{BrainBART-Large} (w/ \textsc{C-SCL}) &\textbf{39.14} &\textbf{20.35} &\textbf{37.12} &\textbf{35.91} &\textbf{25.96} &\textbf{21.31} &18.89 & \textbf{68.48}  \\
            \midrule
            \textsc{BrainBART-Base}  & 36.46 &17.75 &34.23 &33.64 &23.60 &18.78 &16.23 &73.01\\
            \textsc{BrainBART-Base} (w/ \textsc{SCL}) &36.70 &17.92 &34.55 &34.18 &24.07 &19.31 &16.79 & 72.27 \\
            \textsc{BrainBART-Base} (w/ \textsc{C-SCL}) &\textbf{37.01} &\textbf{18.05} &\textbf{34.69} &\textbf{34.55} &\textbf{24.39} &\textbf{19.61} &\textbf{17.04} &\textbf{71.65}   \\
            \midrule
            \midrule
            \textsc{BrainPEGASUS-Large} &37.50 &16.10 &34.27 &34.56 &22.57 &17.07 &14.26 &76.21  \\
            \textsc{BrainPEGASUS-Large} (w/ \textsc{SCL}) &39.34 &18.07 &35.83  &36.35 &24.74 &19.38 &16.62 &74.54 \\
            \textsc{BrainPEGASUS-Large} (w/ \textsc{C-SCL}) &\textbf{40.18} &\textbf{19.20} & \textbf{36.72} &\textbf{37.24} &\textbf{25.89} &\textbf{20.63} &\textbf{17.92} &\textbf{73.43}  \\
            \midrule
            \textsc{BrainPEGASUS-Base} &36.70 &14.37 &33.23 &33.74 & 21.05 &14.80 &11.53 & 78.19 \\
            \textsc{BrainPEGASUS-Base} (w/ \textsc{SCL}) &36.74 &\textbf{15.33} &33.29 &33.84 &\textbf{21.88} &\textbf{16.38} &\textbf{13.60} &77.95  \\
            \textsc{BrainPEGASUS-Base} (w/ \textsc{C-SCL}) &\textbf{37.27} &15.21 &\textbf{33.66} &\textbf{34.20} &21.73 & 16.26 &13.50 & \textbf{76.59} \\
            \midrule
            \midrule
            \textsc{BrainT5-Large} &32.17 &12.12 &29.81 &30.43 &19.24 &13.48 &10.32 &83.69  \\
            \textsc{BrainT5-Large} (w/ \textsc{SCL}) &32.65 &14.84 &30.33 &31.06 &20.80 &15.87 &13.25 &82.61  \\
            \textsc{BrainT5-Large} (w/ \textsc{C-SCL}) &\textbf{32.87} &\textbf{14.87} &\textbf{30.54} &\textbf{31.18} &\textbf{20.91} &\textbf{15.98} &\textbf{13.40} &\textbf{81.91}  \\
            \midrule
            \textsc{BrainT5-Base} &31.12 &7.77 &27.65  &27.05 &13.31 &6.44 &3.38 &86.46 \\
            \textsc{BrainT5-Base} (w/ \textsc{SCL}) &31.37 &8.56 &\textbf{28.17}  &28.38 &14.90 &\textbf{8.08} &4.81 &86.15 \\
            \textsc{BrainT5-Base} (w/ \textsc{C-SCL}) &\textbf{31.38} &\textbf{8.63} &28.15  &\textbf{28.46} &\textbf{14.95} &8.06 &\textbf{4.86} &\textbf{85.10} \\
            \bottomrule
        \end{tabular}
\caption{Test set results on the ZuCo benchmark. ↑ means higher is better. ↓ means lower is better.} \label{tab:main_res}
\end{table*}

\subsection{Backbone Model}
Our backbone model \textsc{BrainTranslator} inherits a typical Encoder-Decoder framework, which first encodes a sequence of word-level EEG features $\Me$ to distributed representations and then generates the target sentence $\CalS$ with the decoder.
Concretely, the model is composed of the \textit{pre-encoder} and the \textit{pre-trained seq2seq model}, in which the pre-encoder serves as a bridge between representation spaces of EEG and text.
Formally speaking, the overall model is formulated as:
\begin{equation}
\small
\begin{split}
 \Me^{N_1} &= \texttt{{Pre-Encoder}}(\Me)\stackrel{N_1}{\underset{n=1}{:=}} \textsc{Ffn}\left(\textsc{Att}(\Me^{n-1})\right)  \\
 \Mx^{{N_1}+{N_2}} &= \texttt{{Pre-trained Encoder}}(\Me^{N_1})\\&\stackrel{N_2}{\underset{n=1}{:=}} \textsc{Ffn}\left(\textsc{Att}(\Mx^{n-1})\right)  \\
  \My^M &= \texttt{{Pre-trained Decoder}}(\My^0,\Mx^{{N_1}+{N_2}})\\
  &\stackrel{M}{\underset{m=1}{:=}} \textsc{Ffn}\left(\textsc{Att}\left(\textsc{Att}(\My^{m-1}), \Mx^{{N_1}+{N_2}}\right)\right)
\end{split}
\label{eq:model}
\end{equation}
where $\stackrel{N}{\underset{n=1}{:=}}$ denotes $N$ identical encoding layers and $\stackrel{M}{\underset{m=1}{:=}}$ denotes $M$ decoding layers.
$\My^0$ describes the shifted right version of $\CalS$, $\textsc{Ffn}(\cdot)$ represents a position-wise feed-forward network, and $\textsc{Att}(\cdot)$ represents a multi-head attention \footnote{Detailed illustration for \textsc{BrainTranslator} is shown in the supplementary file.}.

\subsection{Learning Procedure}
The overall training process follows a two-step manner.

Firstly, we adopt our \textsc{C-SCL} to train the pre-encoder, Algorithm~\ref{algo:cl} shows the entire strategy. 
Formally, we have the contrastive triple $(\Me_i, \Me_i^+, \Me_i^-)$ for a given anchor $\Me_i$.
After the transformation of the pre-encoder, we can get $(\Vh_i, \Vh_i^+, \Vh_i^-)$, where $\Vh_i$ is the averaged vector of the outputs of the pre-encoder.
Following the contrastive framework in \newcite{gao2021simcse}, we minimize the cross-entropy loss $\ell_i$ defined by ($N$ is the mini-batch size):
\begin{equation}
    \label{eq:sup_objective}
    \begin{aligned}
        \ell_i = - \log \frac{e^{{\texttt{sim}}({\Vh}_i,{\Vh}_i^+)/ \tau }}{\sum_{j=1}^N\left(e^{{\texttt{sim}}({\Vh}_i,{\Vh}_i^+)/\tau}+e^{{\texttt{sim}}({\Vh}_i,{\Vh}_i^-)/ \tau}\right)}
    \end{aligned}
\end{equation}
where $\tau$ is a temperature hyperparameter\footnote{The key for successful EEG contrastive training is the tiny $\tau$, we show our parameter search experiments in the supplementary file.}. $\texttt{sim}({\Vh}_i,{\Vh}_j)$ is the cosine similarity. 
Note that our \textsc{SCL} works in an online manner, which means both positive and negative pairs are constructed dynamically along with the training process. This increases the distribution of contrastive pairs, thus improving training efficiency. 

Secondly, based on the contrastive-trained pre-encoder, we jointly fine-tune all the parameters of the \textsc{BrainTranslator} to minimize the cross-entropy loss in a parallel training corpus $({\SetE},{\SetS})$: 
\begin{equation}
    {\CalL} =  {-\sum}_{(\Me, \CalS) \in (\SetE,\SetS)} \log p (\CalS \,|\, \Me; \theta) 
\end{equation}

\section{Experiments}
\subsection{Baseline Models}
We adopt the previous state-of-the-art \textbf{\textsc{BrainBART}} \cite{wang2022open} as our baseline model, which is composed of the Transformer pre-encoder\footnote{We also try Conformer \cite{conformer} as the pre-encoder. However, the experimental results show no major difference. Accordingly, we keep using the Transformer pre-encoder in our paper.} and the BART pre-trained seq2seq model \cite{Lewis2019BARTDS}. 
Besides, we further employ other two types of widely used pre-trained seq2seq models, including PEGASUS \cite{zhang2020pegasus} and T5 \cite{raffel2020exploring}, building upon the Transformer pre-encoder to form \textbf{\textsc{BrainPEGASUS}} and \textbf{\textsc{BrainT5}} respectively.
All the above three models come in two model-size variants, including \textbf{\textsc{Large}} and \textbf{\textsc{Base}}, leading to six models in total.

\subsection{Evaluation Protocol}
Following \newcite{wang2022open}, we adopt \textbf{ROUGE} \cite{lin2004rouge} and \textbf{BLEU} \cite{papineni2002bleu} for evaluating our EEG-to-Text generation task. 
Besides, following \newcite{metzger2022generalizable}, we also adopt \textbf{Word Error Rate (WER)} as our metric to examine more fine-grained generation performance.

\subsection{Implementation Details}
Our pre-encoder consists of 6 layers, each with 8 heads and a hidden dimension of 2048. The dimension of the input EEG representation is 840.
For the contrastive training process, we use Adam with learning rate of 0.001 with a batch size of 32. $\tau$ is set to 0.00001.
For the overall training process, we first load the checkpoint of contrastive-trained pre-encoder and then fine-tune the whole model using Adam with learning rate of 2e-5 and batch size of 32.
For the generation process, following \newcite{wang2022open}, we equip our model with greedy decoding to produce final sentences.
For all three metrics, we use standard implementations provided by HuggingFace\footnote{https://github.com/huggingface/evaluate}.

\section{Results}
\subsection{Automatic Evaluation}
Table \ref{tab:main_res} shows the performance of our \textsc{SCL} and \textsc{C-SCL} on the ZuCo benchmark. 
Overall, we can find that \textsc{SCL} can consistently attain strong performance across various baseline models and architectures. With the enhancement of curriculum learning, \textsc{C-SCL} can further boost performance. Except for the main observations, our empirical results also demonstrate the following two findings.
Firstly, \textsc{BART} performs well. 
Although this finding is exclusively derived from results based on three pre-trained seq2seq models, it still provides the guideline for choosing future backbone seq2seq models for EEG-to-Text generation task: choosing task-agnostic language models (e.g., BART) rather than task-oriented models (e.g., PEGASUS for summarization and T5 requiring task prompts).
Secondly, EEG-to-Text generation also follows the scaling law, which means the generation performance scales up with the increasing number of model parameters.

\subsection{Analysis}

\begin{figure}[t]
    \centering
    \includegraphics[scale=0.29]{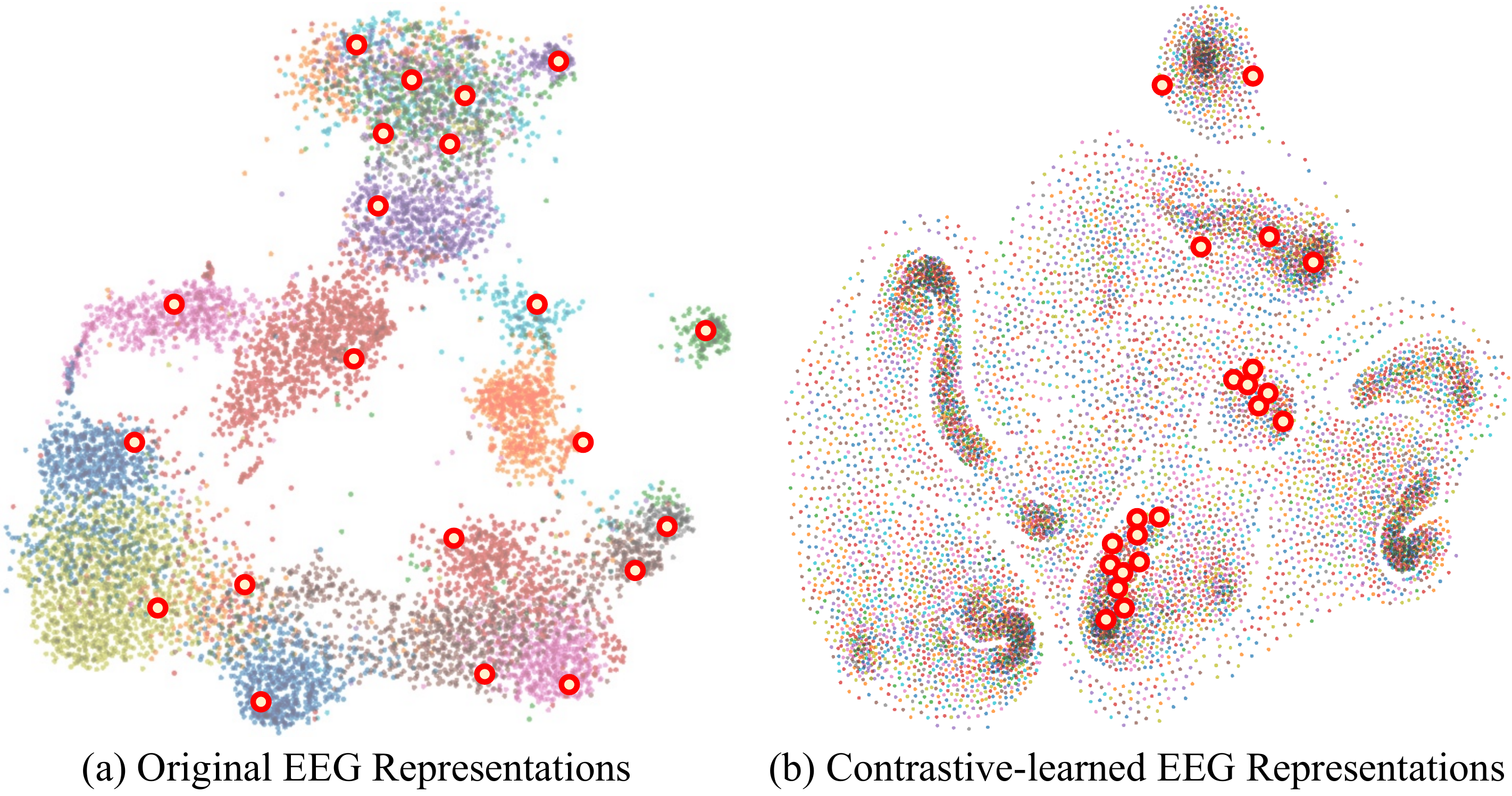}
    \caption{t-SNE visualization of sentence-level EEG representations of sentences in the training set, which are (a) original EEG representations and (b) generated by the pre-encoder after \textsc{C-SCL}. Different colours mean different subjects. Each dot represents a sentence. The red box dots represent the EEG representations corresponding to the same sentence “He and his wife had seven children”.}
    \label{fig:contrastive_embed}
\end{figure}

\noindent \paragraph{Embedding visualization.} 
To verify whether our \textsc{C-SCL} can achieve learning semantic-dependent EEG representations.
We give a straightforward comparison via t-SNE between the original EEG representations (Figure~\ref{fig:contrastive_embed}(a)) and EEG representations obtained after the transformation of the contrastive-trained pre-encoder (Figure~\ref{fig:contrastive_embed}(b)).
We can easily observe that our learned EEG representations of the same sentence tend to be closer compared with original desultorily distributed ones.
This result coincides with our initial goal. 
Besides, Figure~\ref{fig:contrastive_embed}(a) also shows distinct subject clusters (different colours)\footnote{We provide subject labels for each distinct cluster in the supplementary file.} while Figure~\ref{fig:contrastive_embed}(b) reveals the more equally distributed subjects.
Nevertheless, Figure~\ref{fig:contrastive_embed}(b) also shows the EEG representations of the same sentence are not fully clustered. Instead, multiple sub-clusters are formed, which indicates achieving a desirable semantic-dependent EEG representation space is a challenging task.

\begin{figure}[t]
    \centering
    \includegraphics[scale=0.58]{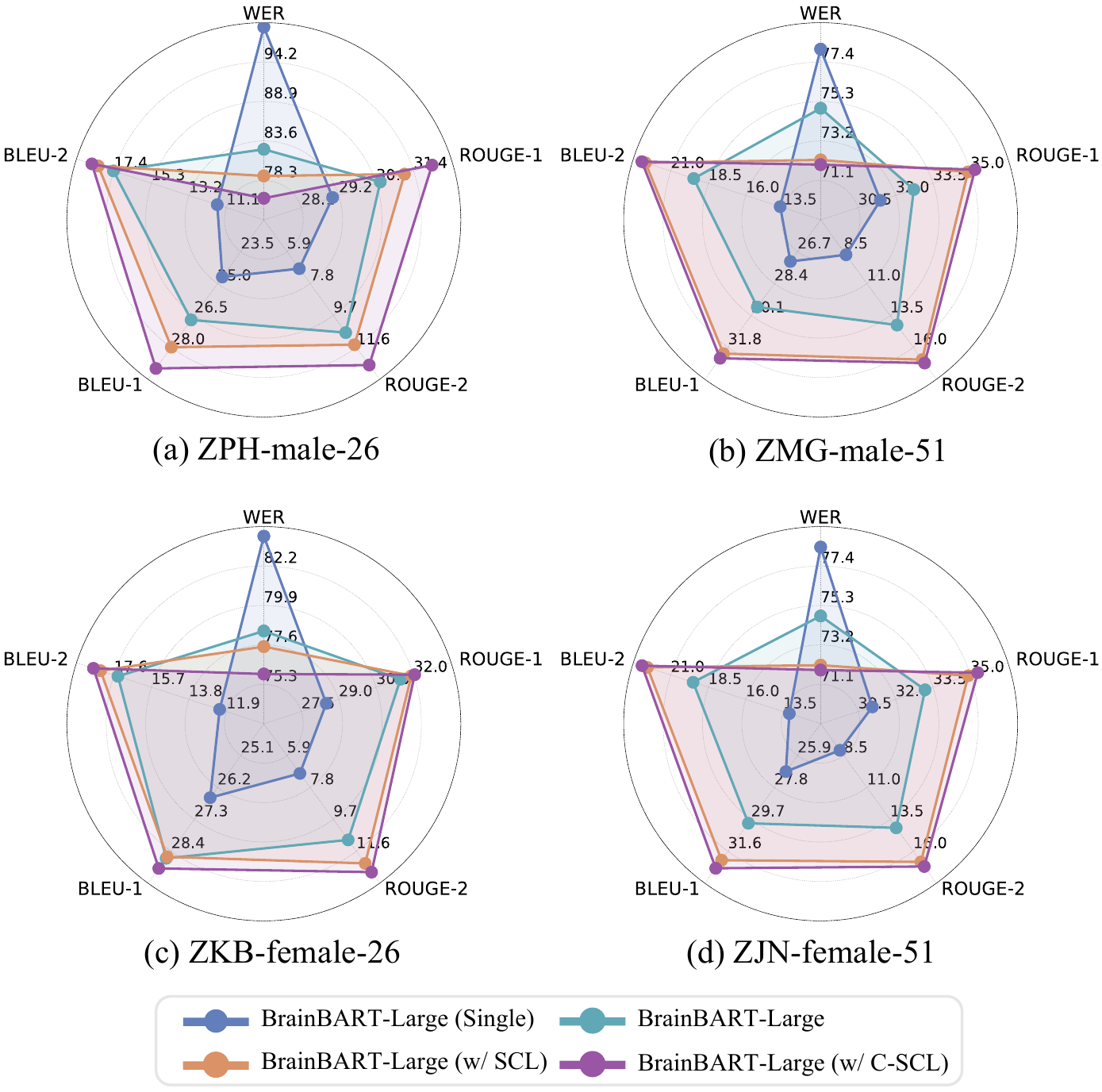}
    \caption{Results of different methods testing on 4 subjects respectively, including both male and female, youth and middle-aged, e.g., ZPH-male-26 describes the subject identified as ZPH, is male and 26 years old. \textsc{BrainBART-Large} (Single) means that training and testing on the data of a single subject. Others mean that training on the whole data while testing on the data of a single subject.}
    \label{fig:radar}
\end{figure}

\noindent \paragraph{Single-subject setting.}
Given that the subject-dependent EEG representation poses a great challenge to the EEG-to-Text generation task, in this analysis, we aim to answer one question: \textit{Whether single-subject training is a more suitable way for the EEG-to-Text generation task?}
To verify this, we test both mixed-subjects training and single-subject training methods on data of 4 distinct subjects.
The results are shown in Figure~\ref{fig:radar}.
Compared with single-subject training, all other three mixed-subjects training methods achieve remarkable improvements, which precisely indicate that it is worth exploring mixed-subjects training methods.
Besides, the results also show the effectiveness of our proposed \textsc{SCL} and \textsc{C-SCL} at a more fine-grained level.

\begin{figure*}[t]
    \centering
    \includegraphics[scale=0.5]{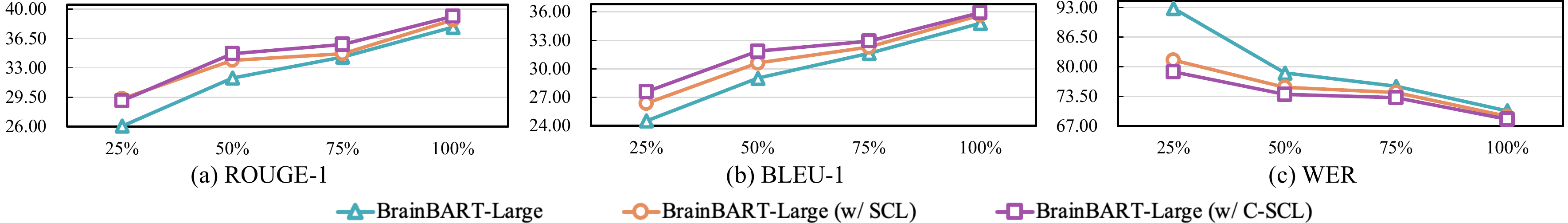}
    \caption{Results of different training data sizes.}
    \label{fig:percent}
\end{figure*}

\noindent \paragraph{Low-resource setting.}
To verify the robustness of our methods on varying data sizes, we provide datasets of different sizes to train the pre-encoder using \textsc{SCL} and \textsc{C-SCL}, then fine-tune the whole model.
Note that the size of the test set is the same across all experiments.
The results are shown in Figure~\ref{fig:percent}.
We can find that the model performance clearly improves with the growing of dataset size in terms of different types of metrics.
Prominently, our methods show great advantages in the low-resource setting.
Especially when only using 25\% of the dataset, our \textsc{C-SCL} can directly reduce the WER from 92.83\% to 78.89\%, achieving comparable results compared with using 50\% of the dataset.

\begin{figure}[t]
    \centering
    \includegraphics[scale=0.47]{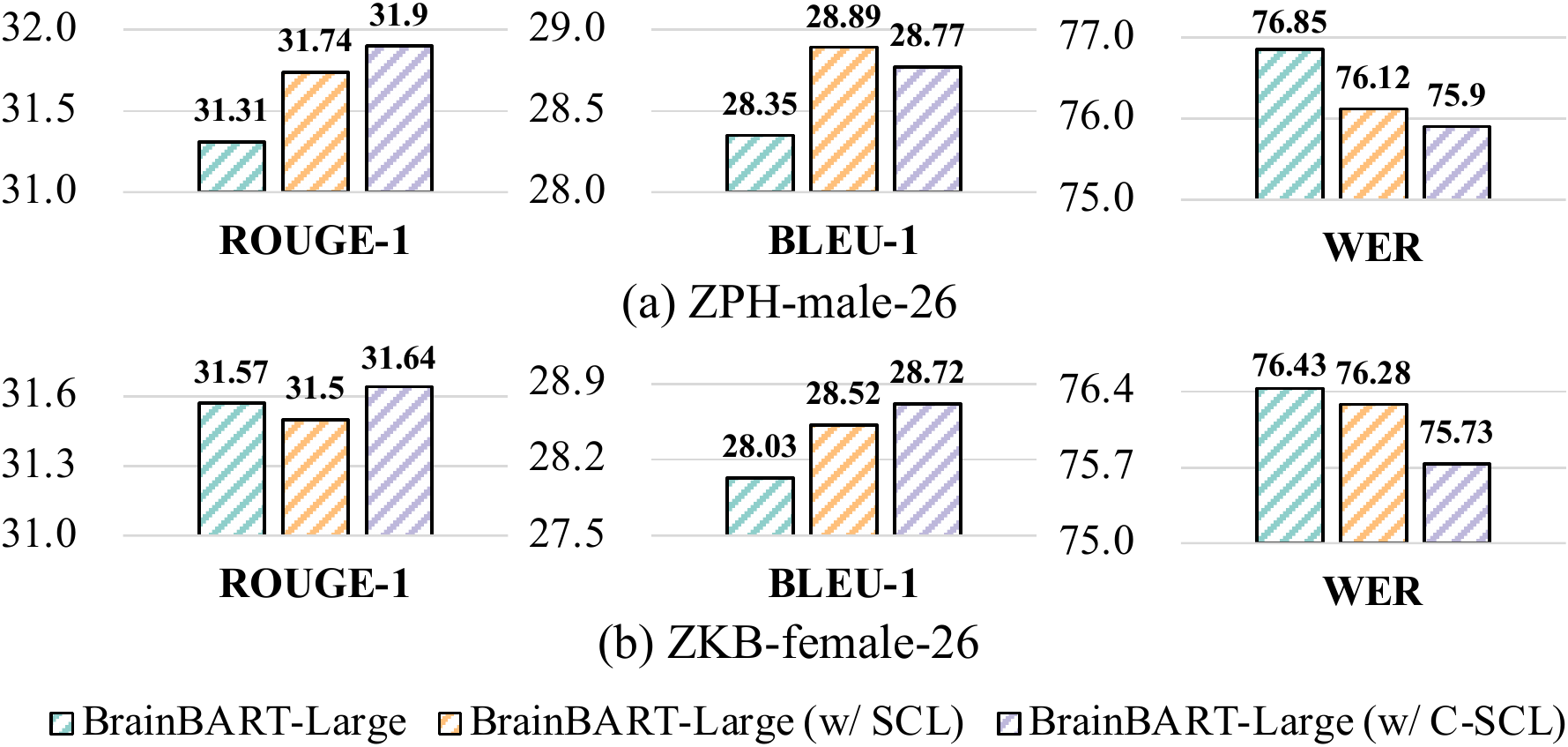}
    \caption{Zero-shot results by training on data that excluded the final test subject.}
    \label{fig:leave_out}
\end{figure}

\noindent \paragraph{Zero-shot setting.}
To verify the generalizability of our methods, we conduct zero-shot experiments by training on the partial ZuCo dataset, which excludes the data of one selected test subject. 
The results are shown in Figure~\ref{fig:leave_out}.
We can see that our methods yield strong performance for unseen ZPH and ZKP respectively.
We attribute this good generalizability to the fact that contrastive learning not only learns better representations for currently available subjects but also optimizes a distinguishable representation space that can be easily transferred and adapted to unseen subjects.

\noindent \paragraph{Single-curriculum setting.}
To verify the necessity of curriculum learning for our \textsc{C-SCL}. 
We individually perform \textsc{SCL} based on contrastive pairs from each curriculum level, including \textsc{SCL}($\SetE^{\texttt{easy}}$), \textsc{SCL}($\SetE^{\texttt{medium}}$) and \textsc{SCL}($\SetE^{\texttt{hard}}$).
Then, we select one third of the data from each curriculum level and conduct \textsc{C-SCL} based on the fixed $[\frac{\SetE^{\texttt{easy}}}{3}, \frac{\SetE^{\texttt{medium}}}{3}, \frac{\SetE^{\texttt{hard}}}{3}]$.
Note that all the above contrastive learning datasets keep the same size and the fine-tuning is based on the whole ZuCo train part.
The results are shown in Table~\ref{tab:single_curr}.
Firstly, we can find that curriculum learning indeed does good to the model performance.
Besides, both \textsc{SCL}($\SetE^{\texttt{easy}}$) and \textsc{SCL}($\SetE^{\texttt{hard}}$) achieve relatively lower results.
We attribute this fact to that easy pairs are insignificant but directly leveraging hard pairs is quite challenging for the model learning.

\begin{table}[!htb]
\centering
        \begin{tabular}{l|lll}
            \toprule
            & \textbf{R-1}↑ & \textbf{B-1}↑ & \textbf{WER}↓ \\
            \midrule
            \textsc{SCL}($\SetE^{\texttt{easy}}$) & 37.89  & 34.82  & 70.05   \\
            \textsc{SCL}($\SetE^{\texttt{medium}}$) & 38.21 & 35.10  & {69.83}   \\
            \textsc{SCL}($\SetE^{\texttt{hard}}$) & 37.92  & 35.08  & 70.09   \\
            \textsc{C-SCL}($[\frac{\SetE^{\texttt{easy}}}{3}, \frac{\SetE^{\texttt{medium}}}{3}, \frac{\SetE^{\texttt{hard}}}{3}]$) & \textbf{38.52}  & \textbf{35.34}  & \textbf{69.49}  \\
            \bottomrule
        \end{tabular}
\caption{Results of different curriculum levels.}
\label{tab:single_curr}
\end{table}

\begin{figure}[t]
    \centering
    \includegraphics[scale=0.84]{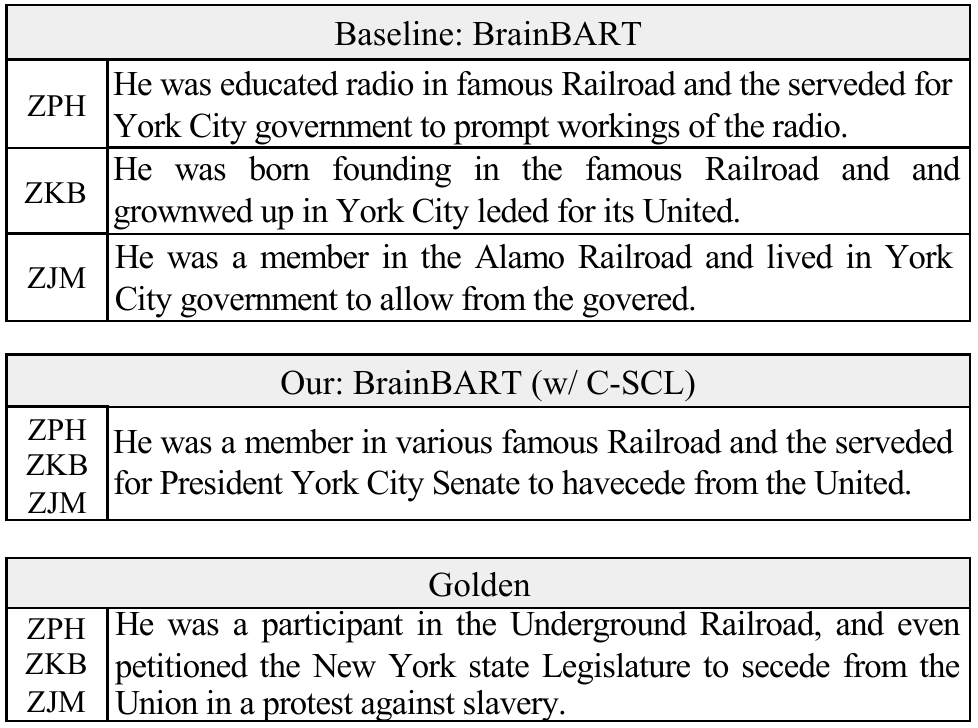}
    \caption{Generations for EEG signals of different subjects. The EEG signals correspond to the same sentence. ``ZPH", ``ZKB", and ``ZJM" are three subjects.}
    \label{fig:case}
\end{figure}

\subsection{Case Study}

Figure~\ref{fig:case} shows the generation case.
We can find that our method can generate the same sentence for EEG signals elicited by different subjects based on learned semantic-dependent EEG representations, whereas the baseline model produces different ones.
Besides, our result is more semantic-related compared with baseline results, which indicates the semantic-dependent EEG representation can enhance the generation performance.
However, there still exists a large gap between our generation and the golden reference. 
We believe future works should pay attention to the following research directions:
(1) Strategies by jointly modelling continuous word-level EEG signals and the syntactic structure of sentences, since current generation still failed to capture the linguistic structure;
(2) Strategies to close the gap between the word-level EEG feature and token-level generation, since current generation still has several spelling errors.

\section{Conclusion}
In this paper, we propose a curriculum semantic-aware contrastive learning strategy (\textsc{C-SCL}) to reduce the discrepancy between the subject-dependent EEG representation and the semantic-dependent text representation.
The experimental results based on the ZuCo benchmark demonstrate its effectiveness for the EEG-to-Text generation task.
Besides, our analyses also verify the robustness and superior generalizability of our \textsc{C-SCL} in the low-resource setting and the zero-shot setting, respectively.
Moreover, single-subject setting experiments also point the necessity of exploring mixed-subjects training methods for the EEG-to-Text generation task.

\bibliographystyle{named}
\bibliography{ijcai23}

\clearpage
\newpage
\appendix

\section*{\textsc{BrainTranslator}}
\begin{figure}[!htb]
    \centering
    \includegraphics[scale=0.62]{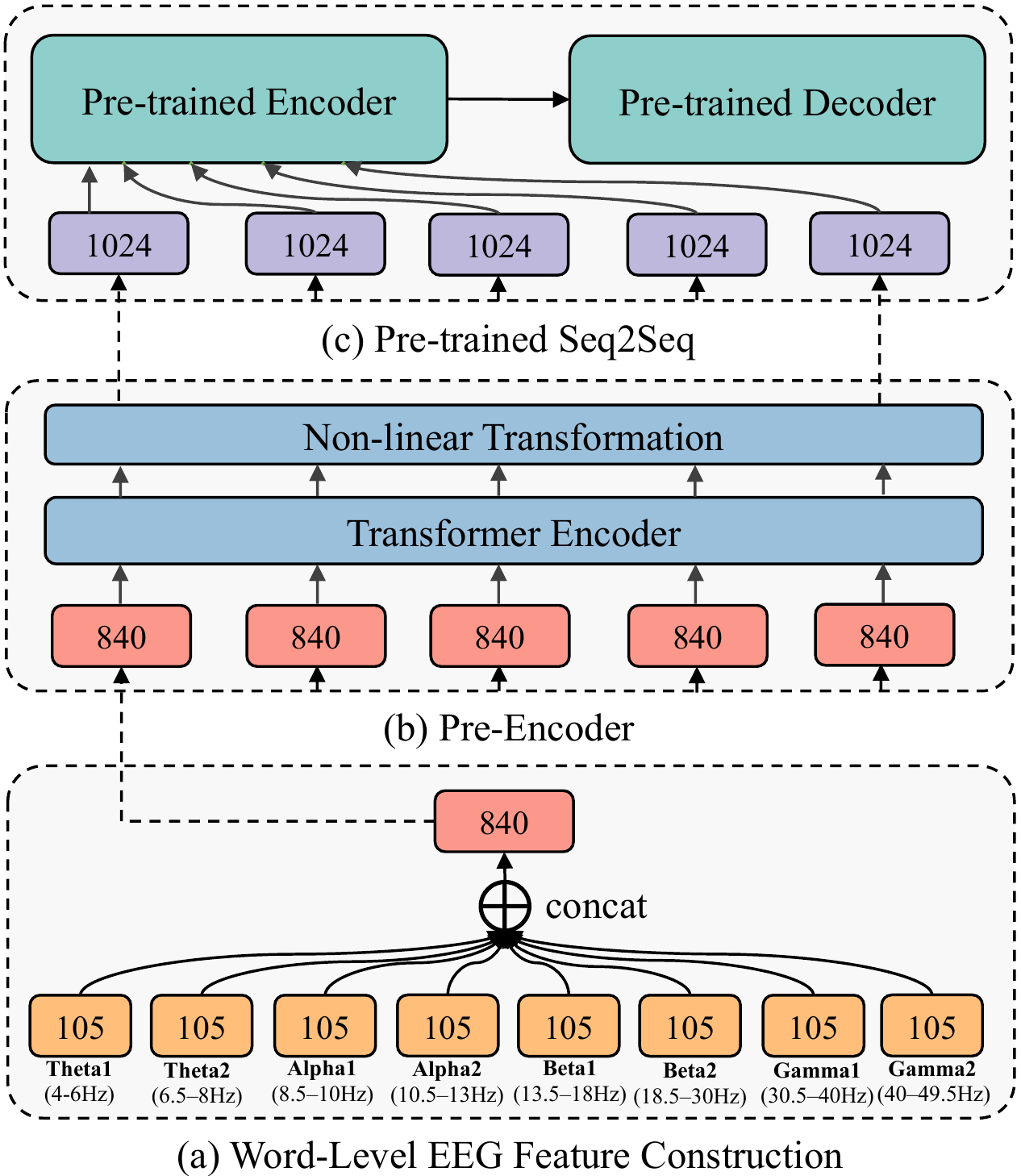}
    \caption{Illustration of the \textsc{BrainTranslator}.}
    \label{fig:model}
\end{figure}

\textsc{BrainTranslator} takes word-level EEG features as input and produces the corresponding sentence. It mainly consists of three parts: (a) Word-Level EEG Feature Construction that concatenates features of different bands of one word to form the final word-level EEG feature. (b) Pre-encoder that transforms original EEG features into the pre-trained Seq2Seq embedding space, and (c) Pre-trained Seq2Seq that takes a sequence of transformed embeddings and produces the final output sentence. The number in the rectangle denotes the dimension of the vector. The overall architecture is shown in Figure~\ref{fig:model}.

\section*{Parameter Search for $\tau$}
\begin{figure}[!htb]
    \centering
    \includegraphics[scale=0.25]{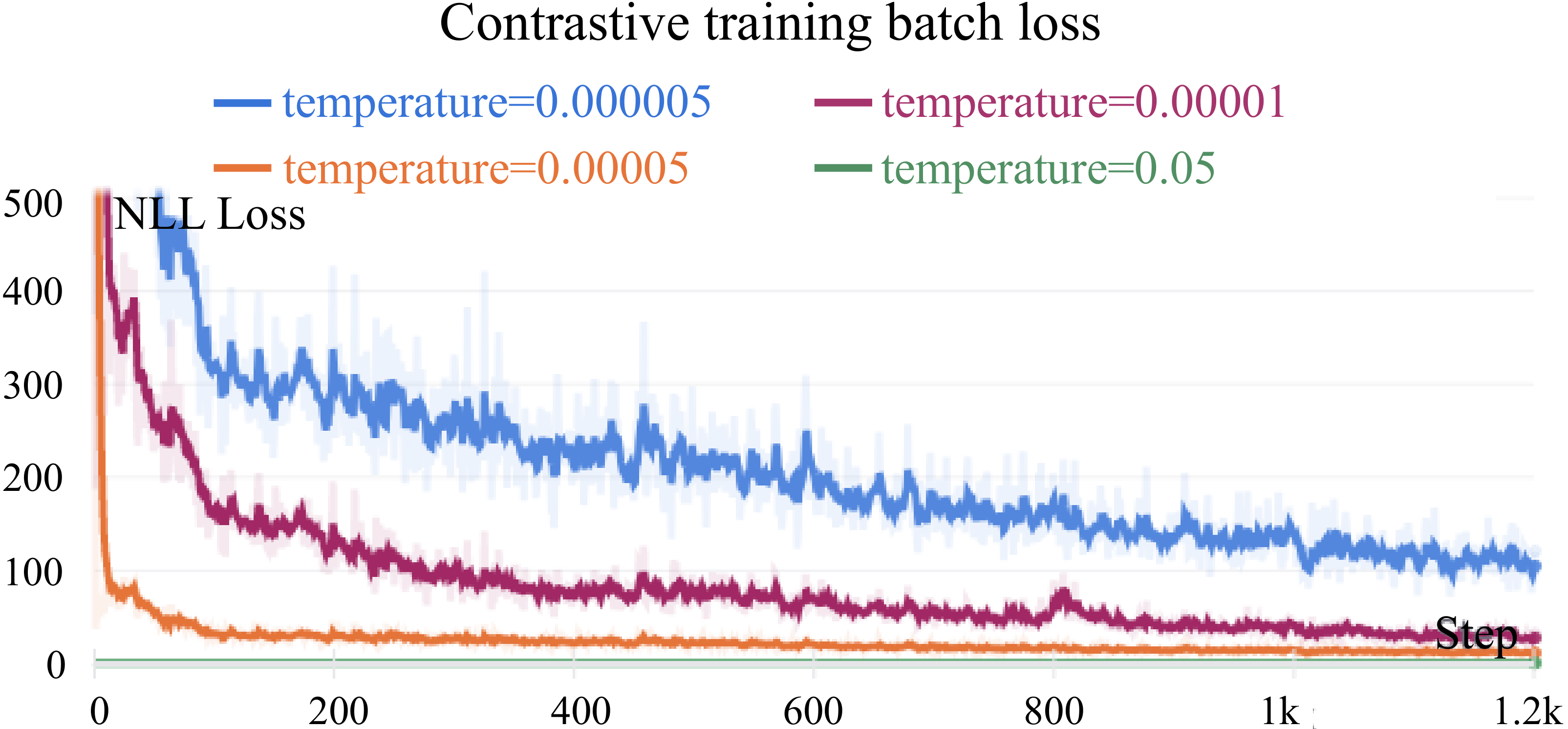}
    \caption{Contrastive training loss with respect to different $\tau$.}
    \label{fig:temp}
\end{figure}

\noindent Figure~\ref{fig:temp} shows the contrastive training loss under different $\tau$\footnote{The figure is obtained via https://wandb.ai/}. 
We can find that setting $\tau$ to a small number is critical for successful EEG contrastive training.
This is due the original EEG signals are similar to each other, a small $\tau$ can produce more distinguishable EEG representations, thus enabling the effective contrastive learning.
We conduct preliminary experiments and find that setting  $\tau$ to 0.00001 yields better EEG-to-Text generation performance.

\section*{Example Contrastive Pairs}

\begin{figure}[!htb]
    \centering
    \includegraphics[scale=0.44]{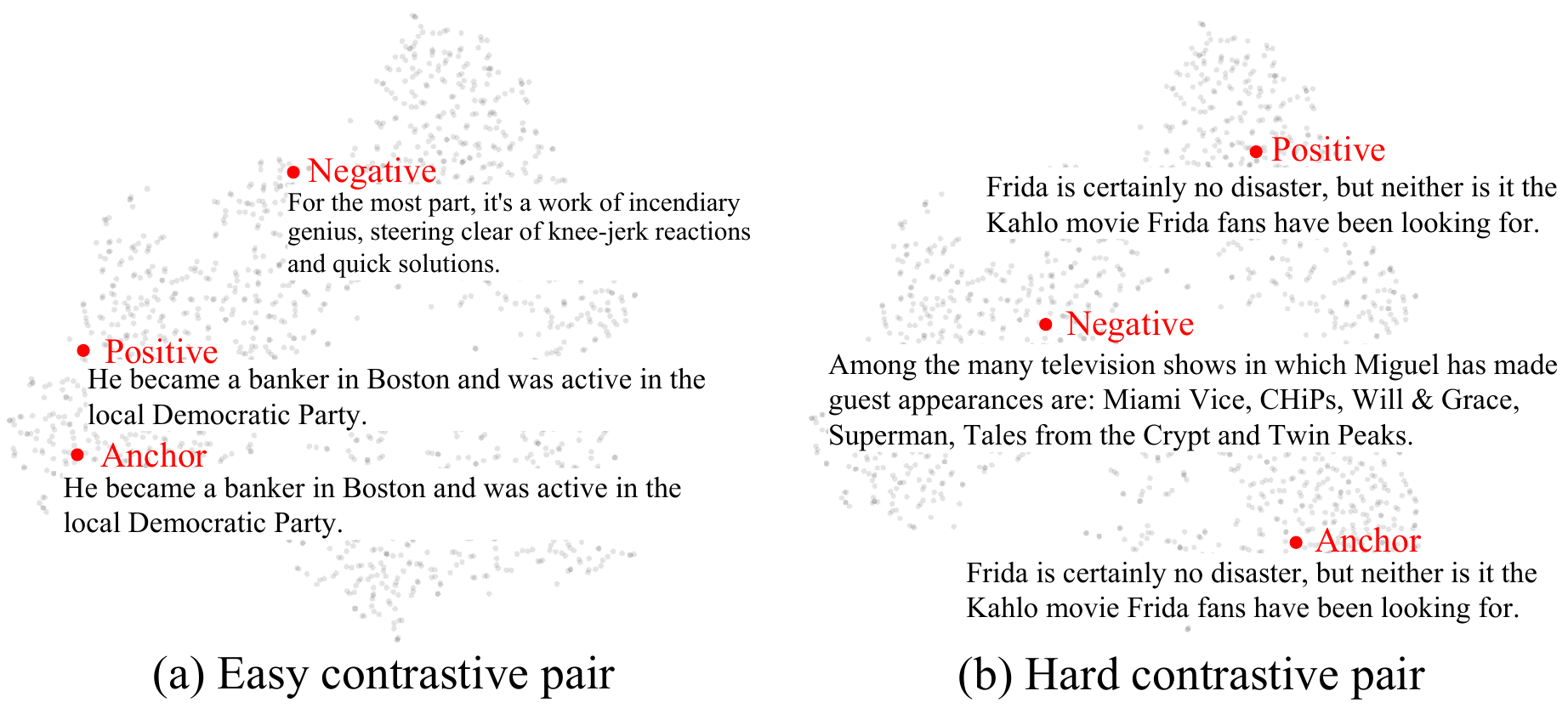}
    \caption{Contrastive pairs of different difficulties.}
    \label{fig:cl_example}
\end{figure}

\noindent Figure~\ref{fig:cl_example} provides two examples for contrastive pairs of different difficulties.
We can clearly find that the easy pair already satisfy the condition: positive pairs are similar while negative pairs are dissimilar.
In contrast, the hard contrastive pair instead follows the condition: positive pairs are dissimilar while negative pairs are similar.

\section*{t-SNE Visualization of Original EEG Signals}

\begin{figure}[!htb]
    \centering
    \includegraphics[scale=0.47]{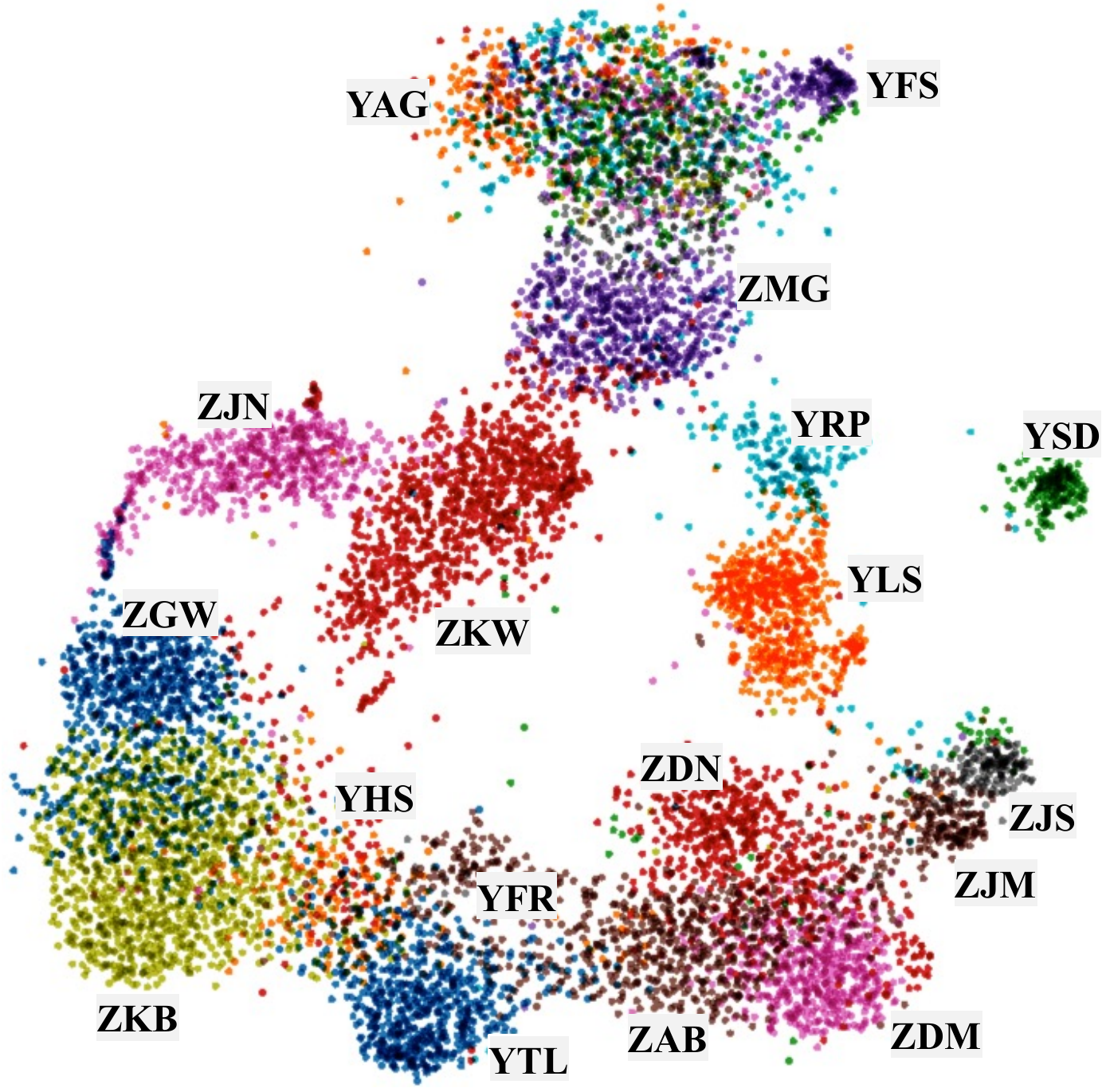}
    \caption{t-SNE visualization of sentence-level EEG representations of sentences in the ZuCo train set. Each dot represents a sentence. Different colours mean different subjects. Distinct clusters are labelled with corresponding subjects.}
    \label{fig:tsne_subject_label}
\end{figure}

\noindent We project sentence-level EEG representation of each sentence in the training set (averaged word-level EEG representations of sentences) to two-dimensional space via t-SNE, as shown in Figure~\ref{fig:tsne_subject_label}.
We can clearly find that the EEG representations elicited by the same subject tend to be closer.

\end{document}